\documentclass[aps,superscriptaddress,amsmath,amssymb,floatfix,twocolumn,showpacs,amsfonts,longbibliography,reprint]{revtex4-2}
\usepackage{times}
\usepackage[varg]{txfonts}
\usepackage{textcomp}
\usepackage{graphicx}
\usepackage{subfigure}
\usepackage{color}
\usepackage[colorlinks=true,citecolor=blue,urlcolor=blue,linkcolor=blue,hyperindex]{hyperref}
\usepackage{braket}
\usepackage{float}
\usepackage{overpic}
\usepackage{bm}
\usepackage{array}
\usepackage{booktabs}
\usepackage{appendix}
\usepackage{tabularx}
\usepackage{cases}
\usepackage{multirow}
\usepackage{mathtools}
\usepackage{soul}

\def\bs{\mathbf{S}}

\begin{document}

\title{RIXS spectra of spinon, doublon, and quarton excitations of a spin-1/2 antiferromagnetic Heisenberg trimer chain}
\author{Junli Li}
\affiliation{State Key Laboratory of Optoelectronic Materials and Technologies, Guangdong Provincial Key Laboratory of Magnetoelectric Physics and Devices, Center for Neutron Science and Technology, School of Physics, Sun Yat-Sen University, Guangzhou 510275, China}
\author{Jun-Qing Cheng}
\affiliation{School of Physical Sciences, Great Bay University, Dongguan 523000, China, and Great Bay Institute for Advanced Study, Dongguan 523000, China}
\author{Trinanjan Datta}
\email[Corresponding author:]{tdatta@augusta.edu}
\affiliation{Department of Physics and Biophysics, Augusta University, 1120 15$^{th}$ Street, Augusta, Georgia 30912, USA}
\affiliation{Kavli Institute for Theoretical Physics, University of California, Santa Barbara, California 93106, USA}
\author{Dao-Xin Yao}
\email[Corresponding author:]{yaodaox@mail.sysu.edu.cn}
\affiliation{State Key Laboratory of Optoelectronic Materials and Technologies, Guangdong Provincial Key Laboratory of Magnetoelectric Physics and Devices, Center for Neutron Science and Technology, School of Physics, Sun Yat-Sen University, Guangzhou 510275, China}
\begin{abstract}
We investigate the excitation spectra of a spin-1/2 antiferromagnetic Heisenberg trimer spin chain by employing a combination of numerical and theoretical techniques. Utilizing the Krylov-space correction-vector method in density matrix renormalization group (DMRG), we calculate both the direct and indirect resonant inelastic x-ray scattering (RIXS) spectra for the trimer spin chain. To interpret the observed features in the RIXS spectra, we perform a theoretical perturbative analysis to compute the energy dispersions which are then utilized to determine the density of states (DOS) for both the single-particle and the two-particle excitation spectra. Our results show that the single-particle continua of the direct RIXS spectrum align with the energy levels observed in the DOS spectra of spinon, doublon, and quarton excitations. Furthermore, the two-particle continua are revealed in the indirect RIXS process, where all possible single particle excitations combine to form the various two-particle excitations of the trimer spin chain. Based on our calculations, we propose the RIXS mechanism of generating the fractionalized (spinon) and collective (doublon and quarton) excitations in the trimer spin chain at both the $L$-edge and the $K$-edge, including discussing the interplay of these excitations in the RIXS spectrum for various trimer coupling strength. The computed energy range of the excitations suggest the possibility of experimental detection at both the $L$-edge and the $K$-edge within the current capabilities of RIXS instrumentation resolution. 
\end{abstract}

\maketitle

\section{Introduction}\label{sec:intro}
Magnon and multimagnon excitations in the square lattice antiferromagnet (AF)~\cite{LuoPhysRevB.89.165103,XiongPhysRevB.96.144436,Huang_2017,HePhysRevB.101.024426} and the triangular lattice AF~\cite{LuoPhysRevB.92.035109,JinPhysRevB.100.054410} have been theoretically investigated using resonant inelastic x-ray scattering (RIXS)~\cite{ament2010strongthreemagnonscatteringcuprates} and Raman scattering~\cite{ChaoPhysRevB.103.024417,LiPhysRevB.107.184402} spectroscopy. However, lattice dimensionality is crucial to the inherent fundamental characteristics of the excitation spectrum harbored by the model and supported in the actual real material. It is well known that the magnon, which would survive in a two- or three-dimensional crystal structure, loses its composite character in one-dimensional (1D) compounds and fractionalizes into spinons and holons ~\cite{Giamarchi10.1093/acprof:oso/9780198525004.001.0001,Kumar_2018,SchlappaNatComm2018,KumarPhysRevB.102.075134,WangPhysRevB.105.184429}.

\begin{figure*}[t]
\centering
\centerline{\includegraphics[width=18.0cm]{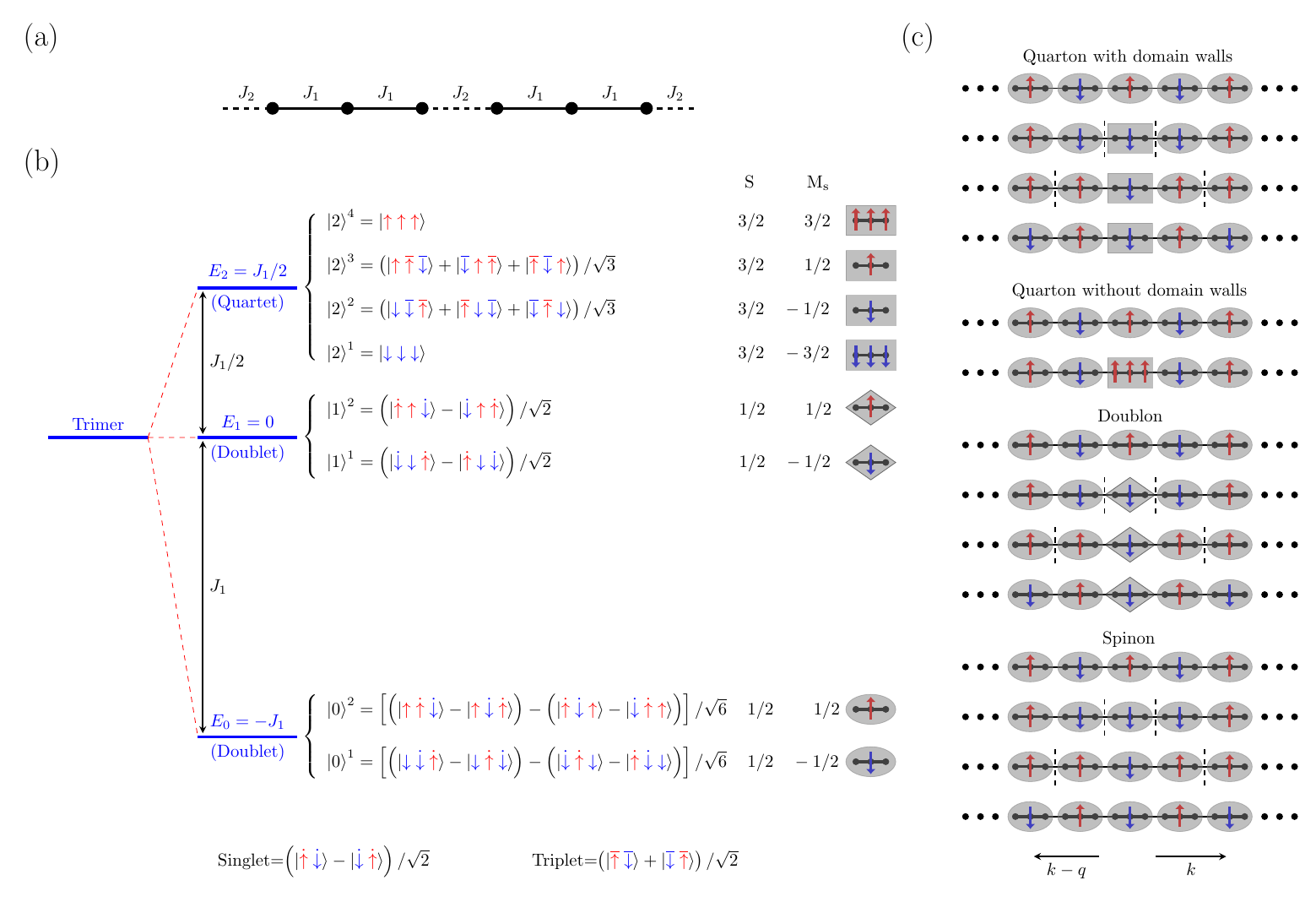}}
\caption{Trimer spin chain interactions and schematic energy level diagram of a single spin trimer unit. (a) A 1D trimer lattice with intratrimer interaction $J_1$ and intertrimer interaction $J_2$ with trimer coupling strength $\mathrm{g}=J_2/J_1$. (b) The possible states for a single trimer where $\ket{0}^{1,2}$, $\ket{1}^{1,2}$ and $\ket{2}^{1,2,3,4}$ refer to the ground state energy level ($E_0=-J_1$ doublet), the intermediate state energy level ($E_1=0$ doublet), and the highest energy level ($E_2=J_1/2$ quartet), respectively. The superscripts on the ket distinguish the degenerate states for each energy level. The total spin quantum number S and the spin projection quantum number M$_\mathrm{s}$ values are shown next to the energy levels. Spin-up (-down) is represented with a red (blue) arrow. The doublet ground state is represented by an ellipse, the excited doublet state is represented by a diamond, and a rectangle represents the excited quartet state. (c) Spinon, doublon and quarton excitations with and without domain wall propagation in a spin trimer chain. For each cartoon we show the ground state configuration $\ket{\psi_G}$ for the entire spin chain. The spinon, the doublon, and the quarton excitation states are specified by $\ket{\psi^n_S}$, $\ket{\psi^n_D}$, and $\ket{\psi^n_Q}$, respectively. The expression for these kets are specified in Sec.~\ref{sec:model}. Dashed lines represent domain walls. The momenta are represented with $k$ and $q$.}
\label{fig:fig1}
\end{figure*}

Recently, the quantum trimer spin system has been investigated both in 1D systems~\cite{MatsudaPhysRevB.71.144411,Drillon199383,HasePhysRevB.73.104419,HasePhysRevB.76.064431,YamamotoPhysRevB.76.014409,HasegawaJPSJ.81.094712,ChengNPJQM2022,BeraNatComm2022,cheng2024} and in higher dimensional lattices~\cite{CaoNPJQM2020,HasegawaJPSJ.81.094712,JanduraPhysRevResearch.2.033382,chang2024magnon}. It has been shown that the antiferromagnetic quantum trimer spin-1/2 chain can support both fractionalized and composite spin excitations~\cite{ChengNPJQM2022}. The basic spin trimer unit is composed of three adjacent sites, as shown in Fig.~\ref{fig:fig1}(a). The Hilbert space is composed of three states -- a doublet ground state, an excited doublet state, and an excited quartet state. A single spin-flip excitation within the ground state manifold of the trimer gives rise to spinons. However, the spin-flip excitation can raise the energy of the ground state to a higher excited state of the trimer unit. In this case, the excited doublet and the quartet states become accessible. The collective excitations which arise from the excited doublet and the quartet quantum states are called doublons and quartons, respectively. The trimer spin chain has an inherent ability to host fractionalized (spinons) and composite (doublons and quartons) excitations~\cite{ChengNPJQM2022}. Thus, the 1D spin chain trimer is a rich physical system where one can study the interplay of exotic fractionalized and composite excitations. From a material perspective, this trimer spin model can be realized in real chemical compounds Na$_2$Cu$_3$Ge$_4$O$_{12}$~\cite{BeraNatComm2022}, Cu$_{3}$(P$_{2}$O$_{6}$OH)$_{2}$~\cite{HasePhysRevB.73.104419}, and Cu$_{3}$(P$_{2}$O$_{6}$OD)$_{2}$~\cite{HasePhysRevB.76.064431}.

The trimer spin chain shown in Fig.~\ref{fig:fig1}(a) has been studied for various different ratio of the intratrimer ($J_1$) and intertrimer ($J_2$) exchange interactions~\cite{ChengNPJQM2022}. We define the trimer coupling strength as $\mathrm{g}=J_2/J_1$. The spinon excitation in the chain is caused by a spin-flip. Such a spin-flip process does not break the singlet state in the trimer, but only flips the unpaired spin as seen in Fig.~\ref{fig:fig1}(b). This also results in two domain walls on both sides of the trimer [see Fig.~\ref{fig:fig1}(c)]. Doublon and quarton excitations also include single spin-flip processes, as seen in Fig.~\ref{fig:fig1}(c). A doublon is created when a spin flips in a trimer with the singlet in the ground state being broken. The spins form a new singlet in the same trimer, creating an excited doublet state. The trimer is energetically raised to the intermediate energy level when a doublon is generated. Quarton excitation occurs when the singlet states in the ground state are all broken and replaced by triplet states. The trimer is raised into a quartet state and occupies the highest energy level. Spinon, doublon, and quarton excitations can be revealed by the dynamical structure factor (DSF) of the 1D trimer spin chain \cite{ChengNPJQM2022}. Specifically, the doublon and the quarton excitations have been observed in the inelastic neutron scattering (INS) measurements on Na$_2$Cu$_3$Ge$_4$O$_{12}$~\cite{BeraNatComm2022}.
 
The excitation spectrum of a single doublon and a single quarton has been studied in the spin chain trimer system using quantum Monte Carlo (QMC), exact diagonalization (ED)~\cite{ChengNPJQM2022}, and  density matrix renormalization group (DMRG)~\cite{BeraNatComm2022,cheng2024}. Trimer spin chains for $\mathrm{g}>1$~\cite{HasePhysRevB.73.104419} and $\mathrm{g}<1$~\cite{BeraNatComm2022,cheng2024} have the same static magnetic response due to the formation of a one-third magnetization plateau under the induction of a magnetic field. In contrast to the $\mathrm{g}<1$ case~\cite{ChengNPJQM2022}, no high energy excitations are observed in the INS spectrum of the $\mathrm{g}>1$ trimer spin chain~\cite{HasePhysRevB.76.064431}. When $\mathrm{g}<1$, QMC and ED computations indicate that the two-spinon continuum splits into a low energy part and a high energy part. The high energy part continues to decompose further for $\mathrm{g}<0.7$~\cite{ChengNPJQM2022}, resulting in an intermediate energy zone and a high energy band. The intermediate energy region arises from the doublon excitation and the high energy band can be attributed to the quarton excitation. When the doublon excitation occurs, a trimer is excited into a doublet excited state with an energy cost of $\omega=E_1-E_0=J_1$. For the quarton excitation, the trimer is excited from a doublet ground state to a quartet excited state, with an energy cost of $\omega=E_2-E_0=3J_1 /2$.  The doublon and quarton excitations with higher energy can be observed evidently if g decreases below 0.4.  In Ref.~\cite{ChengNPJQM2022}, the numerical results from QMC simulations with subsequent numerical analytic continuation of the imaginary-time correlation function suggested that the spinon, the doublon, and the quarton excitations coexist clearly in a trimer spin chain when $\mathrm{g}\leq0.4$. Note, when $\mathrm{g}=1$, the trimer chain transforms into a uniform Heisenberg AF chain, where only the two-spinon continuum is present.

Density matrix renormalization group is a highly efficient and accurate computational paradigm to calculate the ground state spin ordering~\cite{WhitePhysRevLett.69.2863,WhitePhysRevB.48.10345,Feiguin10.1063/1.3667323,ALVAREZ20122226} and spin dynamics~\cite{WhitePhysRevLett.93.076401,Daley_2004,SchollwockRevModPhys.77.259,ALVAREZ20091572,AlvarezPhysRevE.84.056706,Feiguin10.1063/1.3667323,AlvarezPhysRevB.87.245130,KuhnerPhysRevB.60.335,JeckelmannPhysRevB.66.045114,NoceraPhysRevE.94.053308,NoceraPhysRevB.106.205106} of the 1D and the quasi-1D Heisenberg spin systems. We perform Krylov-space correction-vector (CV) method in DMRG simulations to compute both the single-particle and the two-particle  excitation spectrum of the $J_1$-$J_1$-$J_2$ spin-1/2 trimer chain at zero temperature. Specifically, we elucidate the role of excitations in the trimer spin chain that can be probed both by direct and indirect RIXS spectroscopy. Direct RIXS provides a straightforward observation of the DSF where single-particle excitations including the spinon, the doublon, and the quarton excitation can be observed. The indirect RIXS process creates two-particle excitations such as the spinon-doublon, the two-doublon, and the doublon-quarton excitations. Two-particle excitations in a trimer spin chain represents two quasiparticles that are excited at two adjacent trimer, respectively. Inelastic neutron scattering has been the dominant experimental technique to explore magnetic excitations in a quantum trimer spin chain~\cite{MatsudaPhysRevB.71.144411,HasePhysRevB.76.064431}. Resonant inelastic x-ray scattering spectroscopy has been applied to investigate a layered two-dimensional trimer system Ba$_4$Ir$_3$O$_{10}$~\cite{ShenPhysRevLett.129.207201}. With a similar motivation and based on our calculations, we demonstrate that RIXS is a viable spectroscopic method to detect both single-particle and two-particle excitations in a trimer spin chain at both the $L$-edge and the $K$-edge. 

In this article, we utilize a combination of numerical and theoretical techniques to analyze the energy spectrum of the AF trimer spin-1/2 chain. Numerically, we calculate the direct and the indirect RIXS spectrum using the Krylov-space CV method in DMRG simulations. To unravel the features observed in the RIXS spectra, we use a perturbative analysis to compute the energy dispersion which serves as an input to calculate the density of states (DOS) spectra for the single-particle and the two-particle excitations. The single-particle continua of the calculated direct RIXS spectrum agrees with the energy levels observed in the DOS spectra of the spinon, the doublon, and the quarton excitations. The two-particle continua are revealed in the indirect RIXS process. All six possible combinations of single particle excitations of the trimer spin chain can be observed in the indirect RIXS spectrum. The two-particle excitations -- the two-spinon, the spinon-doublon, the spinon-quarton, the two-doublon, the doublon-quarton, and the two-quarton excitation -- span from the low energy region to the high energy region of the indirect RIXS spectrum. Based on our calculations, we elucidate the RIXS mechanism of generating fractionalized and collective excitations in the trimer spin chain at both the $L$-edge and the $K$-edge. The analysis performed in this article provides a deeper insight into the rich Hilbert space spectrum of a trimer spin chain, which can be accessed by x-ray spectroscopy. The predicted energy range of these excitations lie within the current scope of RIXS experimental detection.

This article is organized as follows. In Sec.~\ref{sec:model}, we introduce the model Hamiltonian of a quantum trimer spin chain and derive the spinon, the doublon, and the quarton excitation dispersions. In Sec.~\ref{sec:method}, we outline the various possible RIXS mechanisms and the expressions for both the direct and the indirect RIXS response functions. We also discuss the DMRG numerical method as applied to our spin chain. In Sec.~\ref{sec:results}, we display our results and discuss the physical origins of the direct and the indirect RIXS spectrum features. We compare the RIXS intensity pattern with the DOS calculation results and experimental INS data. In Sec.~\ref{sec:conclusions}, we provide our conclusions. In Appendix.~\ref{app:rateindrixs}, we list the transition rate probability equations for the intertrimer two-particle excitations.

\section{Model}\label{sec:model}

The Hamiltonian of the antiferromagnetic spin-1/2 trimer chain for length $L$ with $N=L/3$ trimers is given by~\cite{ChengNPJQM2022} \begin{equation}
\begin{split}
\label{eq:ham}
\hat{H}=&\sum^N_{n=1}\left\{J_1\left[\hat{\bs}_{3(n-1)+1}\cdot\hat{\bs}_{3(n-1)+2}+\hat{\bs}_{3(n-1)+2}\cdot\hat{\bs}_{3n}\right]\right. \\
&\left.+J_2\hat{\bs}_{3n}\cdot\hat{\bs}_{3n+1}\right\},
\end{split}
\end{equation}
where $\hat{\bs}_j$ is the spin operator at site $j$ that spans over the trimer site indices shown in Eq.~\eqref{eq:ham}. As mentioned previously, the exchange couplings $J_1$ and $J_2$ denote intratrimer and intertrimer antiferromagnetic exchange interactions, respectively. A schematic representation of this model is shown in Fig.~\ref{fig:fig1}(a). The Hamiltonian of a single spin trimer (a chain with only three sites 1, 2, and 3) can be written as $\hat{H}_{123}=\hat{h}_{12}+\hat{h}_{23}$. Here $\hat{h}_{12}=J_1\left(\hat{S}^x\otimes\hat{S}^x\otimes\hat{I}_{2\times 2}+\hat{S}^y\otimes\hat{S}^y\otimes\hat{I}_{2\times 2}+\hat{S}^z\otimes\hat{S}^z\otimes\hat{I}_{2\times 2}\right)$ and $\hat{h}_{23}=J_1\left(\hat{I}_{2\times 2}\otimes\hat{S}^x\otimes\hat{S}^x+\hat{I}_{2\times 2}\otimes\hat{S}^y\otimes\hat{S}^y+\hat{I}_{2\times 2}\otimes\hat{S}^z\otimes\hat{S}^z\right)$, where $\hat{S}^{\alpha}=\sigma^{\alpha}/2$ ($\alpha=x,y,z$) and $\sigma^\alpha$ represent the Pauli matrices. Diagonalization of $\hat{H}_{123}$ results in three distinct energy levels and eight spin states in one spin trimer as shown in Fig.~\ref{fig:fig1}(b).

In a single-particle excitation process, we define the ground state $\ket{\psi_G}$, the spinon excitation state $\ket{\psi^n_S}$, the doublon excitation state $\ket{\psi^n_D}$, and the quarton excitation state $\ket{\psi^n_Q}$ of a trimer spin chain at the $n$-th trimer with $N$ trimers as 
\begin{equation}
\begin{split}
\label{eq:wfn}
\ket{\psi_G}&=\ket{0}^2_1\;\ket{0}^1_2\cdot\cdot\cdot\ket{0}^1_{n-1}\;\ket{0}^2_n\;\;\ket{0}^1_{n+1}\cdot\cdot\cdot\ket{0}^2_{N-1}\;\ket{0}^1_N, \\
\ket{\psi^n_S}&=\ket{0}^2_1\;\ket{0}^1_2\cdot\cdot\cdot\ket{0}^1_{n-1}\;\ket{0}^1_n\;\;\ket{0}^1_{n+1}\cdot\cdot\cdot\ket{0}^2_{N-1}\;\ket{0}^1_N, \\
\ket{\psi^n_D}&=\ket{0}^2_1\;\ket{0}^1_2\cdot\cdot\cdot\ket{0}^1_{n-1}\;\ket{1}^1_n\;\;\ket{0}^1_{n+1}\cdot\cdot\cdot\ket{0}^2_{N-1}\;\ket{0}^1_N, \\
\ket{\psi^n_Q}&=\ket{0}^2_1\;\ket{0}^1_2\cdot\cdot\cdot\ket{0}^1_{n-1}\;\ket{2}^{2,4}_n\ket{0}^1_{n+1}\cdot\cdot\cdot\ket{0}^2_{N-1}\;\ket{0}^1_N.
\end{split}
\end{equation}
The physical meaning of the numerical value inside the kets and the superscripts are defined in the caption of Fig.~\ref{fig:fig1}(b). The subscripts represent the position of the trimer in the spin chain. Symbols $G$, $S$, $D$, and $Q$ represent the ground state, the spinon, the doublon, and the quarton, respectively. In Eq.~\eqref{eq:wfn}, the spinon, the doublon, and the quarton are all assumed to be excited from the trimer ground state $\ket{0}^2$. Earlier numerical investigations using QMC and ED~\cite{ChengNPJQM2022} indicate that the doublon and the quarton excitations manifest in the system only when the trimer coupling strength g is low. 

We calculate the dispersion relations for the doublon and the quarton using a trimer spin chain model with six trimers. The ground state is defined as\begin{equation}\label{eq:wfgs6}
\ket{\psi_G}=\ket{0}^1_1\;\ket{0}^2_2\;\ket{0}^1_3\;\ket{0}^2_4\;\ket{0}^1_5\;\ket{0}^2_6.
\end{equation} The excited states with a spinon, a doublon, and a quarton at the first site is given by \begin{equation}
\label{eq:wfdq6}
\begin{split}
\ket{\psi^1_S}&=\ket{0}^2_1\;\ket{0}^2_2\;\ket{0}^1_3\;\ket{0}^2_4\;\ket{0}^1_5\;\ket{0}^2_6, \\
\ket{\psi^1_D}&=\ket{1}^2_1\;\ket{0}^2_2\;\ket{0}^1_3\;\ket{0}^2_4\;\ket{0}^1_5\;\ket{0}^2_6, \\
\ket{\psi^1_Q}&=\ket{2}^2_1\;\ket{0}^2_2\;\ket{0}^1_3\;\ket{0}^2_4\;\ket{0}^1_5\;\ket{0}^2_6.
\end{split}
\end{equation} Upon Fourier transformation, the excited states can be expressed in momentum space as\begin{equation}
\label{eq:wfq}
\ket{\psi^q_{S,D,Q}}=\frac{1}{\sqrt{6}}\sum^6_{n=1}e^{iqn}\ket{\psi^n_{S,D,Q}},
\end{equation}
where $n$ is the site index and $q$ is the momentum along the chain. The Hamiltonian of the trimer spin chain with six trimers can be written as \begin{equation}\label{eq:Hn}
\hat{H}=\sum^6_{n=1}\hat{H}_n+\sum^{5}_{n=1}\hat{H}_{n,n+1},
\end{equation}
where $\sum^6_{n=1}\hat{H}$ considers intratrimer interactions and $\sum^{5}_{n=1}\hat{H}_{n,n+1}$ contains intertrimer interactions. The dispersion relation $\omega_S(q)$  for the spinon in the trimer spin chain is given by~\cite{CloizeauxPhysRev.128.2131,KargarianPhysRevA.77.032346,ChengNPJQM2022}\begin{equation}
\label{eq:sdispersion}
\omega_S(q)=\frac{\pi J_{eff}}{2}\left|\mathrm{sin}(3q)\right|,
\end{equation}
where $J_{eff}=4J_2/9$ is the effective interaction obtained from the Kadanoff method \cite{KargarianPhysRevA.77.032346}. The dispersion relation $\omega_D(q)$ for the doublon and $\omega_Q(q)$ for the quarton are given by
\begin{equation}
\label{eq:ddispersion}
\begin{split}
\omega_D(q)+E_0&=\bra{\psi^q_D}\hat{H}\ket{\psi^q_D} \\
                        &=\sum^6_{n=1}\bra{\psi^q_D}\hat{H}_n\ket{\psi^q_D}+\sum^{5}_{n=1}\bra{\psi^q_D}\hat{H}_{n,n+1}\ket{\psi^q_D} \\
                        &=E_1+\frac{J_2}{54}\left[-20-4\left(6+\sqrt{3}\right)\mathrm{cos}(3q)\right. \\
                        &-\left(12-5\sqrt{3}\right)\mathrm{cos}(6q)+4\left(2+\sqrt{3}\right)\mathrm{cos}(9q) \\
                        &\left.+\left(16+3\sqrt{3}\right)\mathrm{cos}(12q)+2\left(6+\sqrt{3}\right)\mathrm{cos}(15q)\right], \\ 
\end{split}
\end{equation}
\begin{equation}
\label{eq:qdispersion}
\begin{split}
\omega_Q(q)+E_0&=\bra{\psi^q_Q}\hat{H}\ket{\psi^q_Q} \\
                        &=\sum^6_{n=1}\bra{\psi^q_Q}\hat{H}_n\ket{\psi^q_Q}+\sum^{5}_{n=1}\bra{\psi^q_Q}\hat{H}_{n,n+1}\ket{\psi^q_Q} \\
                        &=E_2+\frac{J_2}{54}\left[-20-4\left(6-\sqrt{2}\right)\mathrm{cos}(3q)\right. \\
                        &-\left(12-7\sqrt{2}\right)\mathrm{cos}(6q)+\left(8+5\sqrt{2}\right)\mathrm{cos}(9q) \\
                        &\left.+\left(16+3\sqrt{2}\right)\mathrm{cos}(12q)+\left(12+\sqrt{2}\right)\mathrm{cos}(15q)\right],
\end{split}
\end{equation}
where $E_0$, $E_1$, and $E_2$ are the energy levels of the ground state, the excited doublet state, and the excited quartet state. In the next section, we will utilize the energy level structure to demonstrate the possibility of detecting fractionalized and collective magnetic excitations in the trimer spin chain using RIXS spectroscopy.

\section{RIXS Mechanism and Numerical Method}\label{sec:method}
In RIXS, an incident photon excites a core electron to induce a transition to a higher excited state. The subsequent electronic transitions between the excited state and the core level are mediated by interactions between the electrons and the core holes can generate both non spin-flip and spin-flip transitions. When the spin-flip transitions disperse through the material, collective magnetic excitations emerge. As discussed in Sec.~\ref{sec:model}, a trimer spin chain can harbor low energy fractionalized spinon excitation, high energy collective excitations (doublon and quarton), and combinations of them (discussed in the results section). In the subsequent paragraphs, we discuss both the direct and the indirect RIXS mechanism that is capable of generating and detecting these excitations. We present a physical picture that illustrates how RIXS has the capability to induce both fractionalized and collective excitations in a 1D spin chain.

\begin{figure}[t]
\centering
\centerline{\includegraphics[width=8.7cm]{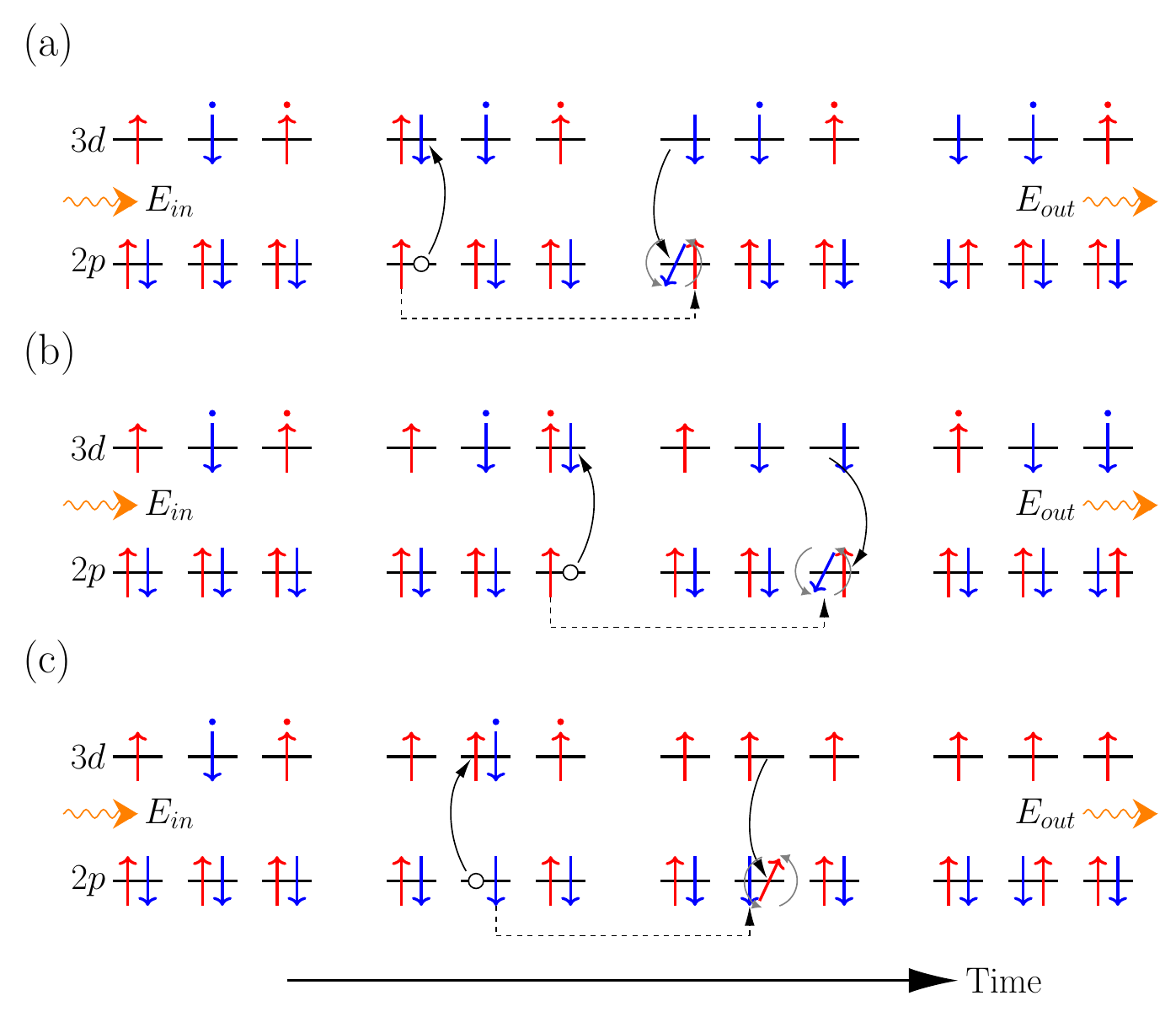}}
\caption{Illustration of direct $L$-edge RIXS process and single-particle excitations of a single trimer unit. Formation of the (a) spinon (b) doublon and (c) quarton excitation. The nearest-neighbor and the next-nearest-neighbor spin pairs with dots represent singlet states.}
\label{fig:fig2}
\end{figure}

In Fig.~\ref{fig:fig2}, we display the direct RIXS mechanism process. For illustrative purposes and to keep the discussion tractable, we consider the spin-down and spin-up components of the ground state $|0\rangle^{1,2}$ doublet wave function. Note, the dotted arrows signify the spin-singlet pair. An incident x-ray photon excites an electron from the 2$p$ band to the 3$d$ valence shell ($L$-edge transition), thereby, creating a core hole in the 2$p$ band. Subsequently, an electron of opposite spin orientation (to preserve Pauli exclusion principle) transits from the 3$d$ valence state to fill the core hole, while a spin-flip (supported by spin-orbit coupling) occurs simultaneously (to preserve Pauli exclusion principle). The final outcome is a spin-flipped state which emits an outgoing photon. From Fig.~\ref{fig:fig2}(a) to Fig.~\ref{fig:fig2}(c), each picture gives one of the possibilities to excite the single particles. In Fig.~\ref{fig:fig2}(a), we outline the process for spinon formation. This new state belongs to $|0\rangle^{1}$, thereby generating spinon excitations. Compared to the initial state, the final state of the trimer can either transit to the ground state with a reversed spin direction or transit into higher excited doublet or quartet states. As a result, three different flavors of excitations -- spinon, doublon, and quarton excitations can be created by the direct RIXS process. When the final state of a trimer contains a nearest neighbor singlet as shown in Fig.~\ref{fig:fig2}(a), the spinon excitation happens. If the final state of a trimer includes a singlet formed from a pair of next-nearest-neighbor spins (on the same trimer unit) as presented in Fig.~\ref{fig:fig2}(b), the doublon excitation is generated. In Fig.~\ref{fig:fig2}(c), the final state with a spin triplet refers to the quarton excitation.

\begin{figure*}[t]
\centerline{\includegraphics[width=17.8cm]{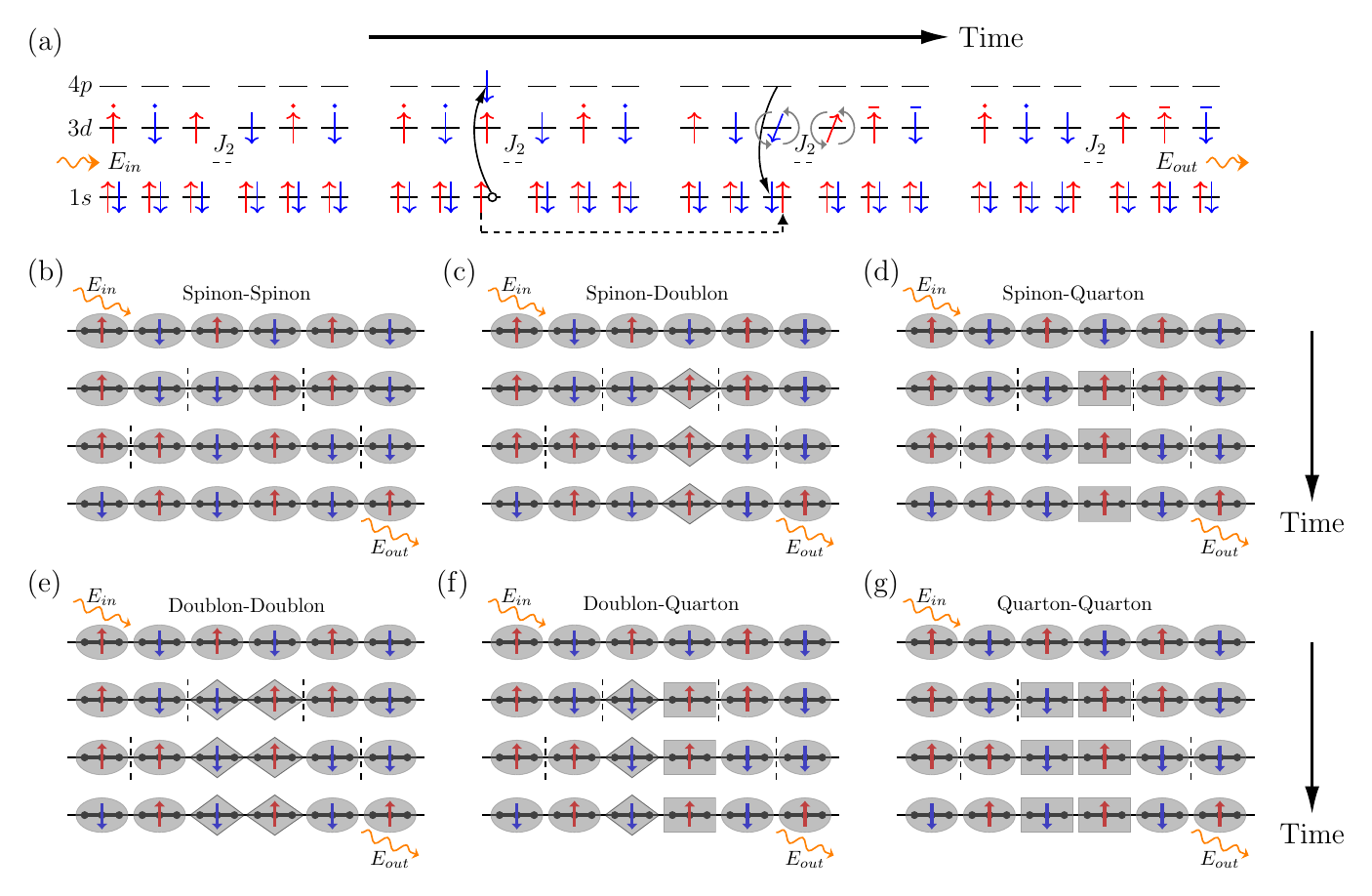}}
\caption{Schematic representation of two-particle excitations which can be detected by $K$-edge RIXS. The ground state with anti-parallel effective spins on nearest-neighbor trimers. (a) Indirect RIXS mechanism of the spinon-quarton excitation. Two-particle RIXS excitation mechanism of the (b) two-spinon (c) spinon-doublon (d) spinon-quarton (e) two-doublon (f) doublon-quarton and (g) two-quarton excitation.}
\label{fig:fig3}
\end{figure*}

In Fig.~\ref{fig:fig3}(a), we illustrate the indirect RIXS mechanism. In this scattering process, we have a  $1s \rightarrow 4p$ ($K$-edge transition) orbital transition. The core hole created by the excited electron transiting to the $4p$ valence shell is scattered and filled by the electron transiting back from the 4$p$ orbital level. A pair of nearest-neighbor spins flip to reverse their original directions, creating a double spin-flip, thus causing the adjacent trimers to be excited into different spin trimer states. This in turn creates collective magnetic excitations in the trimer spin chain. The double spin-flip process generates two-particle excitations which are combinations of the spinon, the doublon, and the quarton. The total number of the different kinds of two-particle excitations are 3!, which is equal to six. These excitation patterns are shown in Figs.~\ref{fig:fig3}(b)-(g). In Fig.~\ref{fig:fig3}(b), the two-spinon excitation occurs when the total spin direction of the two adjacent spin trimers flip. In Fig.~\ref{fig:fig3}(c), the spinon-doublon excitation happens when one spin trimer is excited into the excited doublet state while the other one remains in the ground state. The spinon-quarton excitation is composed of a spin trimer excited into the excited quartet state while the other one stays in the ground state, as shown in Fig.~\ref{fig:fig3}(d). In Figs.~\ref{fig:fig3}(e)-(f), the final state of the system contains one spin trimer at the excited doublet state and the other one is in the excited doublet or quartet state. This refers to the two-doublon or the doublon-quarton excitation, respectively. The two-quarton excitation shown in Fig.~\ref{fig:fig3}(g) contains two trimers excited into the excited quartet states. All the two-particle excitations create two propagating domain walls which contribute to the two-spinon continua signals in the indirect RIXS spectrum.

The direct and the indirect RIXS intensity response functions can be derived analytically based on the UCL expansion~\cite{Brink_2007,FortePhysRevB.77.134428,LuoPhysRevB.92.035109,KumarPhysRevB.99.205130}. When the UCL expansion is applied to express the RIXS intensity functions, the core-hole effect contributes an overall constant in the RIXS scattering amplitude thus making the RIXS intensity proportional to the DSF~\cite{NagNatComm2022}. Therefore, the UCL expansion can be applied to calculate both the direct and the indirect RIXS spectrum. The single spin-flip with $|\Delta\mathrm{M}_\mathrm{s}|=1$ can happen at the Cu $L$-edge ($2p\rightarrow 3d$) in the direct RIXS process, as shown in Fig.~\ref{fig:fig2}. The direct RIXS intensity is given by the expression \begin{equation}
\begin{split}
\label{eq:Sqo}
S(q,\omega)&=\sum_{\alpha=x,y,z}S^{\alpha\alpha}(q,\omega) \\
           &=\sum_{\alpha=x,y,z}\sum_f|\bra{g}\hat{S}^{\alpha}_q\ket{f}|^2\delta(\omega+\omega_g-\omega_f),
\end{split}
\end{equation}
where $\omega$ is the energy, $\ket{g}$ is the ground state, $\ket{f}$ is the final state, and $\omega_f-\omega_g$ is the resonant energy of the direct RIXS process. In a 1D spin system, the DSF should be isotropic in zero magnetic field. Therefore, we have $S^{xx}(q,\omega)=S^{yy}(q,\omega)=S^{zz}(q,\omega)$ in Eq.~\eqref{eq:Sqo}. The single spin-flip form factor is expressed as \begin{equation}
\label{eq:Sq}
\hat{S}^{\alpha}_q=\sum^L_{j=1}e^{iqr_j}\hat{S}^{\alpha}_j\quad,\alpha=x,y,z,
\end{equation}
where $q$ is the momentum and $r_j$ is the position of the $j$-th site. In indirect RIXS, two adjacent spins flip at the same time due to the superexchange interaction. This creates two spinons in the Heisenberg AF chain with g=1.0~\cite{FortePhysRevB.83.245133,KourtisPhysRevB.85.064423}. The indirect RIXS process which can happen at the Cu $K$-edge ($1s\rightarrow 4p$), is shown in Fig~\ref{fig:fig3}. Conservation of angular momentum implies that $|\Delta\mathrm{M}_\mathrm{s}|=0$ for the $K$-edge transition. The indirect RIXS intensity is given by \begin{equation}
\label{eq:Oqo}
O(q,\omega)=\sum_f|\bra{g}O_q\ket{f'}|^2\delta(\omega+\omega_g-\omega_{f'}),
\end{equation}
which contains the two-spin form factor \begin{equation}
\label{eq:Oq}
O_q=\sum^{L-1}_{j=1}e^{iqr_j}\hat{\bs}_j\cdot\hat{\bs}_{j+1}.
\end{equation}
Here $\ket{f'}$ is the final state of the indirect RIXS process and $\omega_{g}-\omega_{f'}$ is the resonant energy.

Density matrix renormalization group algorithm has been used to calculate the DSF of 1D systems~\cite{HallbergPhysRevB.52.R9827,WhitePhysRevLett.93.076401,FeiguinPhysRevB.72.020404,BarthelPhysRevB.79.245101,DargelPhysRevB.85.205119,LakePhysRevLett.111.137205,TakayoshiPhysRevB.99.184303,LaurellPhysRevB.107.104414}. The original DMRG and Lanczos vector method combination gives results with noise, which is only able to show the maximum intensity for the corresponding energy with a lack accuracy~\cite{HallbergPhysRevB.52.R9827,KuhnerPhysRevB.60.335,DargelPhysRevB.85.205119}. The later developed CV method makes tremendous progress at calculating the DSF with higher accuracy~\cite{KuhnerPhysRevB.60.335,JeckelmannPhysRevB.66.045114} to produce results which are able to reach the same resolution as experiments~\cite{NoceraPhysRevB.106.205106,LaurellPhysRevLett.127.037201,NagNatComm2022}. In this article, we utilize the more advanced Krylov-space CV method, which has less error and is computationally more time efficient compared to the original CV method~\cite{NoceraPhysRevE.94.053308}. We implemented these state of the art algorithm in a MATLAB \cite{MATLAB} code to compute the DSF, the direct RIXS spectrum, and the indirect RIXS spectrum at zero temperature using DMRG~\cite{NoceraSciRep2018}. The codes were run in a high performance computing cluster based on a parallel MATLAB paradigm.

The analytical direct and indirect RIXS intensity expressions were recast in the Krylov-space CV representation in order to perform the numerical DMRG calculations. We can write the expressions for the direct RIXS and the indirect RIXS response functions as ~\cite{NoceraPhysRevE.94.053308,NoceraSciRep2018,NagNatComm2022}
\begin{equation}
\begin{split}
\label{eq:Idirect_qo}
S&(q,\omega)=-\frac{1}{\pi}\sum_{\alpha=x,y,z}\sum^L_{j=1}e^{iq\left(r_j-r_c\right)}\times \\
&\mathrm{Im}\left[\bra{g}\hat{S}^{\alpha}_j VU\frac{1}{\omega+i\delta+E_g-\hat{D}}U^{\dagger}V^{\dagger}\hat{S}^{\alpha}_c\ket{g}\right], \\
O&(q,\omega)=-\frac{1}{\pi}\sum_{\alpha,\beta=x,y,z}\sum^{L-1}_{j=1}e^{iq\left(r_j-r_c\right)}\times \\
&\mathrm{Im}\left[\bra{g}\hat{S}^{\alpha}_j\hat{S}^{\alpha}_{j+1}VU\frac{1}{\omega+i\delta+E_g-\hat{D}}U^{\dagger}V^{\dagger}\hat{S}^{\beta}_c\hat{S}^{\beta}_{c+1}\ket{g}\right],
\end{split}
\end{equation}
where $r_c$ is the site position at the center of the lattice. The operator $\hat{D}$ is a diagonal matrix obtained from the Hamiltonian $\hat{H}$ by applying the Lanczos diagonalization method. The $U$ and $V$ are the rotational matrices. The relationship between $U$, $V$ and the diagonalized matrix $\hat{D}$ is given by $U^{\dag}V^{\dag}\hat{H}VU=U^{\dag}\hat{T}U=\hat{D}$, indicating that the diagonal elements in $\hat{D}$ are the eigenvalues of the tridiagonal matrix $\hat{T}$. The tridiagonal matrix $\hat{T}$ is calculated by Lanczos tridiagonalization. In the numerical Krylov-space CV method simulation the parameters are set as follows. The lattice length of the trimer spin chain is $L=60$, the maximum number of kept states is $m=400$ and the truncation error is $10^{-6}$. The maximum Lanczos tridiagonalization step is set to 1000 while keeping the tridiagonalization error below $10^{-7}$. 

We also perform DOS calculations of the trimer spin chain. The DOS intensity functions for the single-particle excitations are given by\begin{equation}
\label{eq:nx}
\begin{split}
n^S(\omega)&=-\frac{1}{\pi}\sum^{2\pi}_{k,q=0}\mathrm{Im}\left[\frac{1}{\omega+i\delta-\omega_S(k-q)-\omega_S(k)}\right], \\
n^D(\omega)&=-\frac{1}{\pi}\sum^{2\pi}_{q=0}\mathrm{Im}\left[\frac{1}{\omega+i\delta-\omega_D(q)}\right], \\
n^Q(\omega)&=-\frac{1}{\pi}\sum^{2\pi}_{q=0}\mathrm{Im}\left[\frac{1}{\omega+i\delta-\omega_Q(q)}\right],
\end{split}
\end{equation}
where the symbols $S$, $D$ and $Q$ have the same meaning as before. The DOS functions for the two-particle excitations are given by \begin{equation}
\label{eq:nxx}
\begin{split}
n^{\alpha\beta}(\omega)&=-\frac{1}{\pi}\sum_{\alpha,\beta\in\left\{S,D,Q\right\}}\sum^{2\pi}_{k,q=0}\times \\
                       &\mathrm{Im}\left[\frac{1}{\omega+i\delta-\omega_\alpha(k-q)-\omega_\beta(k)}\right].
\end{split}
\end{equation}
Although the DOS functions $n^S(\omega)$ and $n^{SS}(\omega)$ are the same, $n^S(\omega)$ contains the spinon excitation while $n^{SS}(\omega)$ is from the two-spinon excitation. All the CV simulation results and the DOS calculations were performed with a spectral broadening $\delta=0.1J_1$.

The intensity of the RIXS signals for various excitations are related to the transition probabilities. For direct RIXS, we assume that the spinon, the doublon, and the quarton are excited from the ground state $\ket{0}^{i}_n$. The transition probabilities for the spinon, the doublon, and the quarton excitation are denoted by $I_{G\rightarrow G}$, $I_{G\rightarrow D}$, and $I_{G\rightarrow Q}$, respectively. These intensities are computed using the expressions 
\begin{subequations}
\begin{eqnarray}
\label{eq:rate1s}
I_{G\rightarrow G}&=&\sum_{\sigma=\pm}\sum_{\left(i,\overline{i}\right)}\sum_{n'=1,2,3}\left|\prescript{\overline{i}}{n}{\bra{0}}S^{\sigma}_{n'}\ket{0}^{i}_n\right|^2, \\
\label{eq:rate1d}
I_{G\rightarrow D}&=&\sum_{\sigma=\pm}\sum_{\left(i,\overline{i}\right)}\sum_{n'=1,2,3}\left|\prescript{\overline{i}}{n}{\bra{1}}S^{\sigma}_{n'}\ket{0}^{i}_n\right|^2, \\
\label{eq:rate1q}
I_{G\rightarrow Q}&=&\sum_{\sigma=\pm}\sum_{\left(i,\overline{j}\right)}\sum_{n'=1,2,3}\left|\prescript{\overline{j}}{n}{\bra{2}}S^{\sigma}_{n'}\ket{0}^{i}_n\right|^2,
\end{eqnarray}
\end{subequations}
where the combinations for $i,\overline{i},j$, and $\overline{j}$ are $(i,\overline{i})\in\left\{(1,2),(2,1)\right\}$, $(\overline{i},i)\in\left\{(1,2),(2,1)\right\}$, $(i,\overline{j})\in\left\{(1,3),(2,2)\right\}$ and $(\overline{i},j)\in\left\{(1,3),(2,2)\right\}$. 

For indirect RIXS, the double spin-flip can occur either in the same trimer or in two adjacent trimers. When the double spin-flip occurs in the same trimer, the ground state is assumed to be $\ket{0}^{i}_n$ and the intensity expressions are given by
\begin{subequations}
\begin{eqnarray}
\label{eq:selects}
I_{G\rightarrow G}&=&\sum_{\sigma=\pm}\sum_{\left(i,\overline{i}\right)}\sum_{n'=1,2}\left|\prescript{\overline{i}}{n}{\bra{0}}S^{\sigma}_{n',n'+1}\ket{0}^{i}_n\right|^2, \\
\label{eq:selectd}
I_{G\rightarrow D}&=&\sum_{\sigma=\pm}\sum_{\left(i,\overline{i}\right)}\sum_{n'=1,2}\left|\prescript{\overline{i}}{n}{\bra{1}}S^{\sigma}_{n',n'+1}\ket{0}^{i}_n\right|^2, \\
\label{eq:selectq}
I_{G\rightarrow Q}&=&\sum_{\sigma=\pm}\sum_{\left(i,\overline{j}\right)}\sum_{n'=1,2}\left|\prescript{\overline{j}}{n}{\bra{2}}S^{\sigma}_{n',n'+1}\ket{0}^{i}_n\right|^2.
\end{eqnarray}
\end{subequations}
When the flip occurs in two adjacent trimers, the ground state is assumed to be $\ket{0}^{i}_n\ket{0}^{\overline{i}}_{n+1}$. The transition possibilities for the two-spinon, the spinon-doublon, the spinon-quarton, the two-doublon, the doublon-quarton, and the two-quarton excitations are denoted by $I_{GG\rightarrow GG}$, $I_{GG\rightarrow GD}$, $I_{GG\rightarrow GQ}$, $I_{GG\rightarrow DD}$, $I_{GG\rightarrow DQ}$, and $I_{GG\rightarrow QQ}$, respectively. To ensure a favorable reading experience these transition probability expressions are supplied in Appendix~\ref{app:rateindrixs}. In Eqs.~\eqref{eq:rate1s}~--~\eqref{eq:rate1q}, Eqs.~\eqref{eq:selects}~--~\eqref{eq:selectq}, and Eqs.~\eqref{eq:rate2ass}~--~\eqref{eq:rate2aqq}, we define $S^{\sigma}_1=S^{\sigma}\otimes I_{2\times 2}\otimes I_{2\times 2}$, $S^{\sigma}_2=I_{2\times 2}\otimes S^{\sigma}\otimes I_{2\times 2}$, $S^{\sigma}_3=I_{2\times 2}\otimes I_{2\times 2}\otimes S^{\sigma}$, $S^{\sigma}_{1,2}=S^{\sigma}\otimes S^{\overline{\sigma}}\otimes I_{2\times 2}$, and $S^{\sigma}_{2,3}=I_{2\times 2}\otimes S^{\sigma}\otimes S^{\overline{\sigma}}$. The relationship between $\sigma$ and $\overline{\sigma}$ is $\left(\sigma,\overline{\sigma}\right)\in\left\{(+,-),(-,+)\right\}$. Results from these equations are tabulated in Table.~\ref{table:1}. The three intratrimer transition rate formulae in Eqs.~\eqref{eq:selects}~--~\eqref{eq:selectq} result in zero intensity. This implies no intratrimer double spin-flip excitation can occur. Having established the basic equations that are needed to calculate the RIXS spectrum, we will compute the $L$-edge and the $K$-edge RIXS spectra using DMRG. These results are presented and discussed in the following section.

\begin{figure*}[t]
\centering
\centerline{\includegraphics[width=18.0cm]{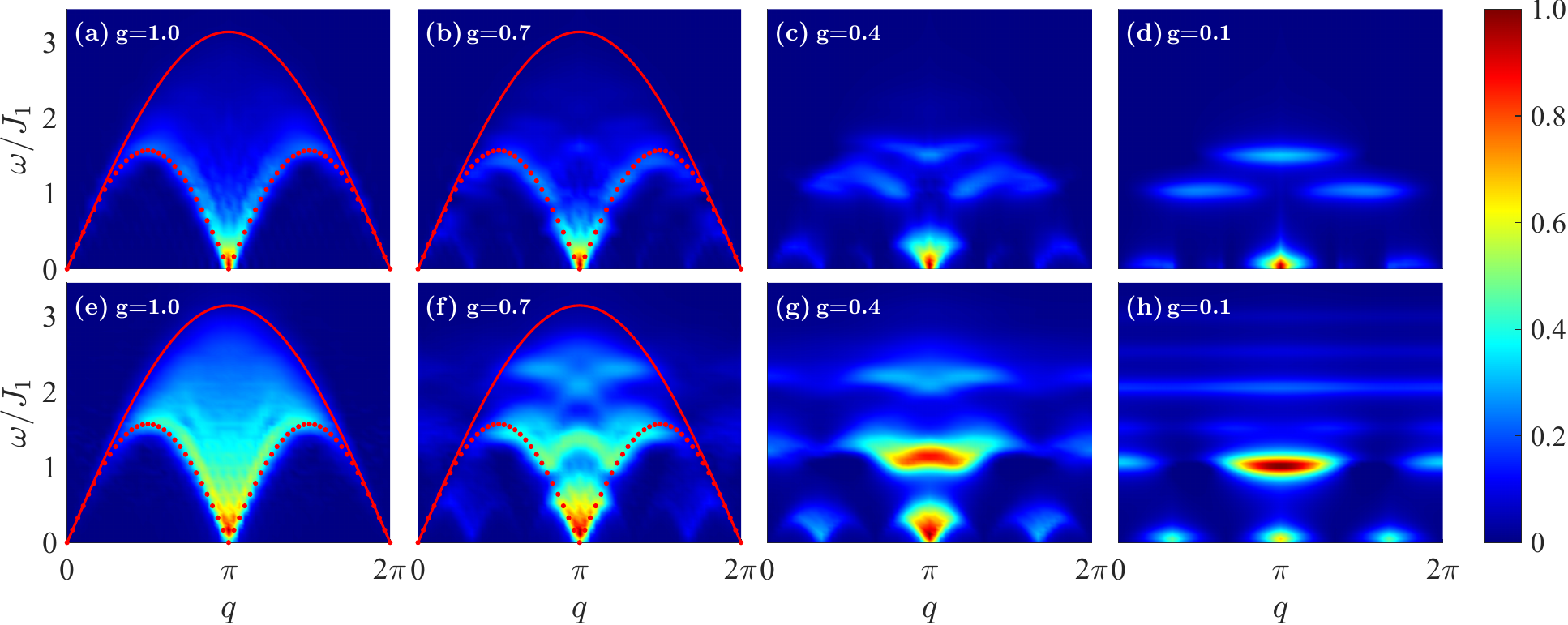}}
\caption{Direct and indirect RIXS spectrum for g$\in$(0,1]. (a)-(d) Direct RIXS spectrum. (e)-(h) Indirect RIXS spectrum. The upper boundary $\omega_U=\pi J_1\left|\sin(q/2)\right|$ and the lower boundary $\omega_L=\pi J_1\left|\sin(q)\right|/2$ of the two-spinon continua for the AF spin-1/2 chain at g = 1.0 are indicated by solid red lines and red dots.}
\label{fig:fig4}
\end{figure*}

\section{Results}\label{sec:results}
In this section, we show the results of the direct RIXS response function $S(q,\omega)$ and the indirect RIXS response function $O(q,\omega)$. The calculations were performed for g$\in$(0,1]. We compare the energy scale of the RIXS spectrum features with the collective magnetic excitation energy to indicate the presence of single-particle and two-particle excitations in both direct and indirect RIXS.

The direct RIXS spectra are shown in Figs.~\ref{fig:fig4}(a)-(d). In Fig.~\ref{fig:fig4}(a), the two-spinon continua for the Heisenberg AF spin chain with the upper boundary $\omega_U =\pi J_1\left|\sin(q/2)\right|$ and the lower boundary $\omega_L=\pi J_1\left|\sin(q)\right|/2$ is sketched with red solid line and red dots, respectively. The maximum energy $J_1$ of the two-spinon continua is reached at $q=\pi$, and both the upper and the lower boundaries touch each other at the zero energy point. The gapless $q=\pi$ point has the highest intensity of the spectral weight while the weaker signals spread throughout the entire two-spinon continua zone. Note, the reproduction of this standard result serves as a quality control check on the implementation of our in-house DMRG code.

In Fig.~\ref{fig:fig4}(b), the two-spinon continua becomes concentrated to lower energy levels. This occurs because the decreased g lowers the average exchange strength of the entire system. Therefore, the spectrum intensity near the upper boundary shrinks and the lower boundary goes below the energy level of the boundary $\omega_L$. At this energy scale, the spin trimer starts to form in the system since $J_2<J_1$ for $\mathrm{g}=0.7$. The spinon propagating between the trimer creates two-spinon continua with lower energy. This formation can be seen as tiny domes of low intensity. The spin trimer with three sites contracts the Brillouin zone (BZ) into $q\in[0,2\pi/3]$ while the BZ of the spin-1/2 Heisenberg AF chain is $q\in[0,2\pi]$. Low energy two-spinon continua structures emerge in the new BZ regions $q\in[0,2\pi/3]$ and $q\in[4\pi/3,2\pi]$. The upper boundary of these low energy two-spinon continua spectrum detach from the two-spinon continua of the Heisenberg AF chain at approximately $\omega\approx J_1$. This signals the onset of the formation of high energy collective excitations that are harbored in the trimer spin chain.

In Fig.~\ref{fig:fig4}(c), when $\mathrm{g}=0.4$, an energy gap opens at $\omega=J_1$. This separates the two-spinon continua into a high energy part and a low energy part. The signals of the upper boundary of the high energy excitation emerges around $\omega=1.5J_1$, which is the same as the quarton energy. Due to the gap opening, the signals of the lower boundary shift to the energy level around $\omega=J_1$, which is equivalent to the doublon energy $J_1$. It can be deduced that when $\mathrm{g}=0.4$, the intratrimer interaction becomes relatively strong enough to form the excited doublet and quartet states, which result in doublon and quarton excitations, respectively. 
Additionally, the low energy spinon excitation has gapless modes at $q=\pm\pi/3$. These modes originate from spinon propagation between each trimer. When the intertrimer coupling strength g is low enough, the antiferromagnetic spin chain disintegrates, and trimers start to form. The original two-spinon upper boundary $\omega_U=\pi J_1|\sin\left(q/2\right)|$ and the lower boundary $\omega_L=\pi J_1|\sin\left(q\right)|/2$ are transformed into $\omega^{\prime}_U=\pi J_{eff}|\sin\left(3q/2\right)|$ and $\omega^{\prime}_L=\pi J_{eff}|\sin\left(3q\right)|/2$, respectively, where $J_1>J_{eff}$. Note, the presence of weak intertrimer interaction reduces the average coupling of the trimer spin chain. This transforms $J_1$ to $J_{eff}$. Also, the momentum transforms from $q$ to $3q$ as the BZ is enlarged due to a transition from one-site (in the antiferromagnetic chain) to three-sites (in a trimer spin chain). Thus, inspecting the expressions for $\omega^{\prime}_U$ and $\omega^{\prime}_L$, we find that the gapless modes show up at both $q=\pi/3$ and $q=-\pi/3$.

We note that for g $\leq$ 0.4, a substantial portion of the spectral weight is localized in the two-spinon continua in the BZ region $q\in[2\pi/3,4\pi/3]$. This can be explained from two aspects. On one hand, the spinon excitation causes zero energy changes for a spin trimer while the doublon and the quarton excitations are high energy excitations localized in a spin trimer. The excitation that absorbs the least energy occurs with the highest probability, which is calculated in Eqs.~\eqref{eq:rate1s}~--~\eqref{eq:rate1q} with the results given in Table.~\ref{table:1}. Hence, the spectrum is more prominent in the low energy region instead of the high energy region. On the other hand, the two-spinon continua in the BZ regions $q\in[0,2\pi/3]$ and $q\in[4\pi/3,2\pi]$ only refer to intertrimer spinon propagation while the one at $q\in[2\pi/3,4\pi/3]$ is the overlap of the intertrimer and intratrimer spinon propagations. Therefore, we notice the presence of most of the spectral weight in the two-spinon continua at $q\in[2\pi/3,4\pi/3]$.


\begin{table*}
\begin{center}
\caption{Transition rates for direct RIXS computed using Eqs.~\eqref{eq:rate1s}~--~\eqref{eq:rate1q} and indirect RIXS calculated using Eqs.~\eqref{eq:rate2ass}~--~\eqref{eq:rate2aqq}. The transition processes are represented by the following symbols: $G\rightarrow G$ (spinon), $G\rightarrow D$ (doublon), $G\rightarrow Q$ (quarton), $GG\rightarrow GG$ (two-spinon), $GG\rightarrow GD$ (spinon-doublon), $GG\rightarrow DD$ (two-doublon), $GG\rightarrow DQ$ (doublon-quarton), and $GG\rightarrow QQ$ (two-quarton). In the second row, we state the transition rate values of the corresponding single-particle and two-particle excitations.} 
\label{table:1}
\begin{tabular}{p{0.15\linewidth}m{0.07\linewidth}<{\centering}m{0.07\linewidth}<{\centering}m{0.07\linewidth}<{\centering}m{0.09\linewidth}<{\centering}m{0.09\linewidth}<{\centering}m{0.09\linewidth}<{\centering}m{0.09\linewidth}<{\centering}m{0.09\linewidth}<{\centering}m{0.09\linewidth}<{\centering}}
\toprule[0.3mm]
\toprule[0.3mm]
Transition process & \textbf{$G\rightarrow G$} & \textbf{$G\rightarrow D$} & \textbf{$G\rightarrow Q$} & \textbf{$GG\rightarrow GG$} & \textbf{$GG\rightarrow GD$} & \textbf{$GG\rightarrow GQ$} & \textbf{$GG\rightarrow DD$} & \textbf{$GG\rightarrow DQ$} & \textbf{$GG\rightarrow QQ$} \\ 
\midrule[0.1mm]
Transition rate & 2.000 & 1.333 & 0.667 & 0.395 & 0.593 & 0.099 & 0.222 & 0.074 & 0.006  \\
\bottomrule[0.2mm]
\bottomrule[0.2mm] 
\end{tabular}   
\end{center}   
\end{table*}

\begin{figure*}[t]
\centering
\centerline{\includegraphics[width=17.8cm]{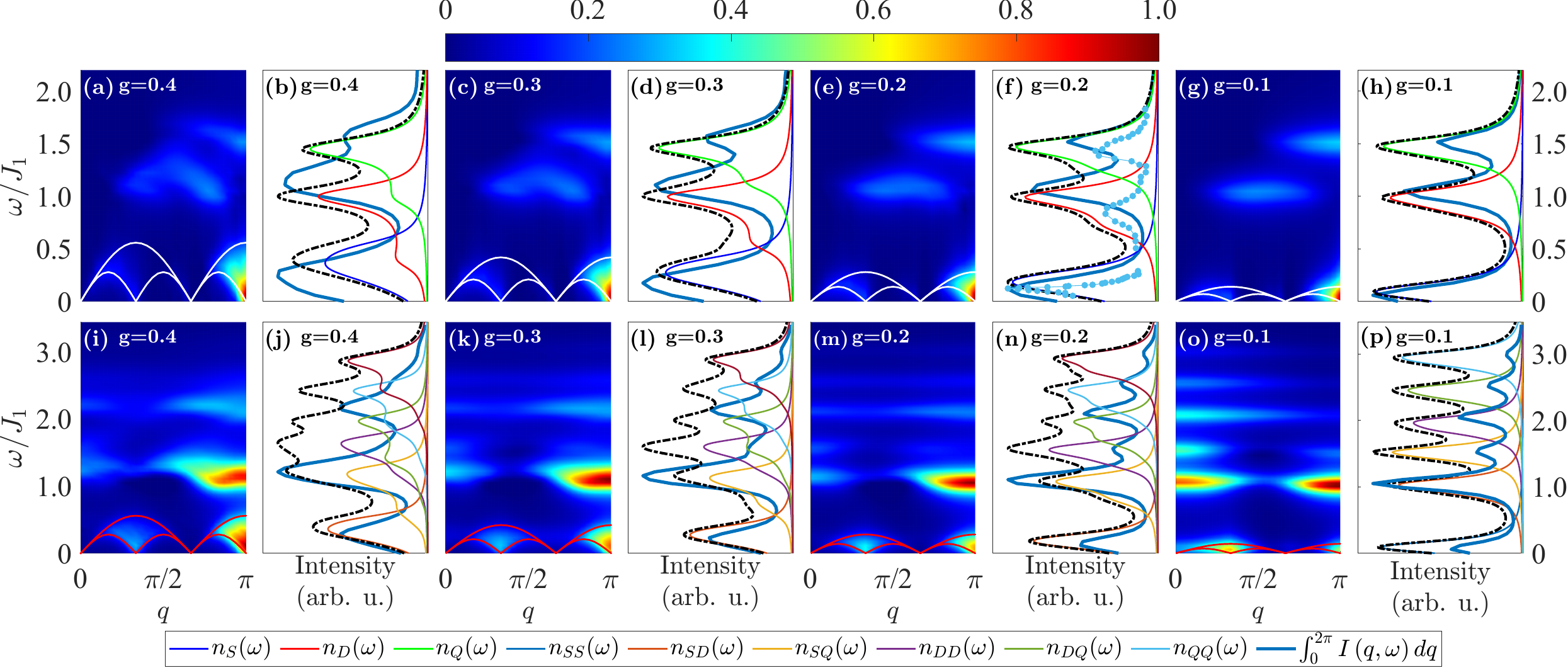}}
\caption{Direct and indirect RIXS spectrum and DOS for single-particle and two-particle excitations. Panels (a),(c),(e), and (g): direct RIXS spectrum. Panels (b),(d),(f), and (h): DOS of single-particle excitations. Panels (i),(k),(m), and (o): indirect RIXS spectrum. (j),(l),(n), and (p): DOS of two-particle excitations. Black dashed curves represent the summations of the DOS spectra in each panel. In the upper panel and the lower panel, we display boundaries of the two-spinon continua in a trimer spin chain by white and red solid lines. For both the white and red solid lines, the upper and lower boundaries are given by $\omega^{\prime}_U=\pi J_{eff}\left|\mathrm{sin}\left(3q/2\right)\right|$ and $\omega^{\prime}_L=\pi J_{eff}\left|\mathrm{sin}\left(3q\right)\right|/2$~\cite{ChengNPJQM2022}, respectively. Whereas in panel (a) a single spin flip excitation is the source of the two-spinon continua (see Fig.~\ref{fig:fig1}(c) for a cartoon description), in panel (i) the two-spinon excitation continua is generated from a double spin-flip in two adjacent trimers (see Fig.~\ref{fig:fig3}(b) for a cartoon description). The blue dots with the solid line in Fig.~\ref{fig:fig5}(f) are INS experimental data for Na$_2$Cu$_3$Ge$_4$O$_{12}$ extracted from Ref.~\cite{BeraNatComm2022} using WebPlotDigitizer~\cite{WebPlotDigitizer}. The blue lines are the integrated RIXS intensities obtained by evaluating $I\left(q,\omega\right)$ over the momentum $q \in [0,2\pi]$ in the BZ, where $I\left(q,\omega\right)$ represents either $S\left(q,\omega\right)$ or $O\left(q,\omega\right)$.}
\label{fig:fig5}
\end{figure*}

In Fig.~\ref{fig:fig4}(d), the upper boundary of the lower energy excitation is subdued to an even lower energy level as g is reduced to 0.1. The high energy continua evolves into flat band structures at two different energy levels, leaving the upper and the lower boundary of the high energy excitation continua precisely at the energy levels $\omega=1.5J_1$ (quarton energy) and $\omega=J_1$ (doublon energy). The weak intertrimer interaction causes the quasiparticles to become localized, resulting in the doublon and the quarton energy bands become flattened. The excitation continua from the low energy and the high energy regions arise from the spinon, the doublon, and the quarton excitations~\cite{ChengNPJQM2022}.

The indirect RIXS spectra are shown in Figs.~\ref{fig:fig4}(e)-(h). In Fig.~\ref{fig:fig4}(e), the upper and the lower boundaries of the two-spinon continua are the same as in Fig.~\ref{fig:fig4}(a), since both the plots are tracking the two-spinon excitation. However, the spectral signals spreading across the continua are caused by a higher resonant energy of the double spin-flip compared to the single spin-flip. In Fig.~\ref{fig:fig4}(f), the two-spinon continua splits into three dominant parts of energy continua. The energy continua at the highest energy level above $\omega=2J_1$ is contributed by the combination of high energy two-particle excitations since the two-particle excitations involving the low energy spinon could not be above the energy of $\omega=1.5J_1$. The energy continua at the intermediate energy level is from the combination of the spinon excitation and the high energy excitations. The intermediate energy continua excludes the pure spinon-spinon excitation for the weak spectral weight around the energy level of $\omega=J_1$. The gapless two-spinon continua structures for intertrimer propagation can be observed at the BZ of $q\in[0,2\pi/3]$ and $q\in[4\pi/3,2\pi]$. The spectral weight in the two-spinon continua is higher than that in the high energy continua, which indicates that the low energy spinon excitation has a greater chance to occur compared to the high energy excitation as calculated in Eqs.~\eqref{eq:rate1s}~--~\eqref{eq:rate1q}. The presence of the high energy continua validates that the separation of the two-spinon continua in Fig.~\ref{fig:fig4}(b) is triggered by the high energy excitations, which are weaker than the low energy spinon excitation.

In Fig.~\ref{fig:fig4}(g), the three energy continua are diffused and localized compared to the ones in Fig.~\ref{fig:fig4}(f), but the primary signatures do not change. However, the intermediate energy continua including both the high energy excitations and the spinon excitation possesses a spectral weight as strong as the spectral weight of the two-spinon continua. Since the reduction in g creates the excited doublet and the quartet states, this in turn enhances the high energy excitation signal. This causes the intermediate energy continua to become relatively strong as the signal in the low energy gapless two-spinon continua which only includes the pure spinon-spinon excitation.

In Fig.~\ref{fig:fig4}(h), the single-particle and the two-particle excitations are visible only when g is decreased to as low as 0.1. It can be observed that six excitation continua spread across the energy region from $\omega=0$ to $\omega=3J_1$. The six excitation continua contributions arise from the two-particle excitations, which are the combinations of spinon, doublon, and quarton excitations. The spectral weights remain mainly at the lowest and the second lowest energy continua. The signals of the higher energy continua are all relatively weaker. The weak energy signals involve the quarton excitation, because the transition probability between the ground state and the excited doublet state is higher compared to the probability of the transition between the ground state and the excited quartet state. The transition rates of the two-particle excitations were evaluated using Eqs.~\eqref{eq:rate2ass}~--~\eqref{eq:rate2aqq}. The results are stated in Table.~\ref{table:1}. It can be inferred from Eqs.~\eqref{eq:rate1s}~--~\eqref{eq:rate1q} that the spinon and the doublon spectral weights occur with the highest strength. This suggests that the trimer spin system has a greater tendency to create spinon-doublon pairs instead of the other two-particle pairs during an indirect RIXS process.

We compare the RIXS spectrum and the DOS (for $\mathrm{g}=0.1,0.2,0.3$, and $0.4$) to distinguish all the single-particle and the two-particle excitation contributions shown in Fig.~\ref{fig:fig5}. The upper row in Fig.~\ref{fig:fig5} are results for direct RIXS. In Fig.~\ref{fig:fig5}(a), we show the direct RIXS spectrum for $\mathrm{g}=0.4$. The two-spinon continua induced by the spinon excitation is surrounded by white lines. The high energy continua between the energy level of $\omega=J_1$ and $\omega=1.5J_1$ are contributed from the broadened DOS spectrum for the doublon and the quarton as shown in Fig.~\ref{fig:fig5}(b). In Fig.~\ref{fig:fig5}(b), the signals for high energy excitations show plateau features which are a consequence of the gradual detachment that occurs when g tends to zero. The plateau for the doublon ranges from $\omega=0.5J_1$ to $\omega=J_1$, while the plateau for the quarton ranges between $\omega=J_1$ to $\omega=1.5J_1$. In Fig.~\ref{fig:fig5}(c), the spectrum for high energy continua are more localized for $\mathrm{g}=0.3$, compared to the high energy continua in Fig.~\ref{fig:fig5}(a). The doublon and the quarton single-particle DOS in Fig.~\ref{fig:fig5}(b) and Fig.~\ref{fig:fig5}(d) display a pronounced a plateau (shoulder) region next to the peaks. As the trimer coupling strength is reduced, this shoulder width shrinks and the DOS weight is upshifted to higher energy. The gradual reshuffling of the DOS weight is further evident as g $\rightarrow$ 0, where eventually there are only peaks and no shoulders. In addition to the above, there is a further rearrangement of the DOS weight. In Fig.~\ref{fig:fig5}(d), the energy level of the DOS signal for the spinon decreases, resulting in a downshift of the upper boundary of the two-spinon continua to a lower energy level around $\omega=0.5J_1$ [see Fig.~\ref{fig:fig5}(c)]. In Fig.~\ref{fig:fig5}(e), the signals for the doublon and the quarton becomes more distinguished and the upper boundary for the two-spinon continua decreases towards a energy level lower than $\omega=0.5J_1$ when $\mathrm{g}=0.2$. In Fig.~\ref{fig:fig5}(f), the plateau for the quarton almost disappears while the plateau in the DOS signal for the doublon has a sliver of intensity around the shoulder region. 

In Fig.~\ref{fig:fig5}(f), we compare the INS experimental data for Na$_2$Cu$_3$Ge$_4$O$_{12}$ from Ref.~\cite{BeraNatComm2022} with our computed DOS spectra for nearest-neighbor interaction at trimer strength $\mathrm{g}=0.2$. Note, the theoretical and experimental analysis in Ref.~\cite{BeraNatComm2022} considers both the nearest-neighbor and the next-nearest-neighbor interaction in Na$_2$Cu$_3$Ge$_4$O$_{12}$. Even though our model is slightly different (due to the neglect of the next-nearest-neighbor interaction), comparing the DOS plots and the INS experimental data suggests that the overall qualitative agreement is good with the slight differences arising in that our model neglects of the next-nearest-neighbor interaction. Both our calculations and Ref.~\cite{BeraNatComm2022} suggest the occurence of the doublon and the quarton excitations. Under the effect of various exchange interactions, the energy levels of the spinon and the doublon excited in Na$_2$Cu$_3$Ge$_4$O$_{12}$ have lower resonance energy compared to the spinon and doublon in our model for $\mathrm{g}=0.2$. But, the quartons from our model and the Na$_2$Cu$_3$Ge$_4$O$_{12}$ have the same excitation energy. The three major signals of the INS data almost aligns with the three DOS spectra signals for the spinon, the doublon, and the quarton excitations. This confirms that both INS and direct RIXS detect the same types of elementary excitations. 

In Fig.~\ref{fig:fig5}(g), the high energy continua disappears with g=0.1, leaving behind distinct upper and lower boundaries that arise from the quarton and the doublon excitations, respectively. The upper boundary of the spinon excitation shifts to an even lower energy level. In Fig.~\ref{fig:fig5}(h), the plateau disappears from the DOS spectrum. Although the DOS intensity for the spinon, the doublon, and the quarton are comparable, the RIXS intensity for the spinon is the highest, which is confirmed by the transition rate calculations in Eqs.~\eqref{eq:rate2ass}~--~\eqref{eq:rate2aqq}. 


The lower row in Fig.~\ref{fig:fig5} presents the results for indirect RIXS. In Fig.~\ref{fig:fig5}(i), the indirect RIXS spectrum in the half BZ region establishes three energy continua at the lowest, the intermediate, and the highest energy levels when g=0.4. that the lowest energy continua is contributed by the two-spinon excitation while the intermediate energy continua is contributed by both spinon-doublon and spinon-quarton excitations. The energy continua at the highest energy level is contributed by two-doublon and doublon-quarton excitations. However, the DOS spectrum in Fig.~\ref{fig:fig5}(j) gives the two-quarton excitation signal at an even higher energy level, which is not visible in the indirect RIXS spectrum in Fig.~\ref{fig:fig5}(i). This reveals that the indirect RIXS is not sensitive to the excitation at the high energy region when the trimer coupling strength is as high as g=0.4. In Fig.~\ref{fig:fig5}(j), the DOS spectrum for spinon-doublon, two-doublon, doublon-quarton, and two-quarton excitations establish plateau structures as seen in Fig.~\ref{fig:fig5}(a).

In Fig.~\ref{fig:fig5}(k), the energy level of the upper boundary for the two-spinon continua decreases while the intermediate and the highest energy continua become localized when g=0.3. There are weak signals at two different energy levels which show up. One is at the energy level above the intermediate energy continua while the other is above the energy level of the highest energy continua. In Fig.~\ref{fig:fig5}(l), energy ranges of the plateau structures become narrow. Fig.~\ref{fig:fig5}(l) indicates that the intermediate and the highest energy continua in Fig.~\ref{fig:fig5}(k) are contributed by the spinon-doublon and the two-doublon excitations, respectively. The weak signals above the intermediate and the highest energy continua refer to the spinon-quarton and the doublon-quarton excitations. In Fig.~\ref{fig:fig5}(m), the upper boundary for the two-spinon continua shifts to an even lower energy level and all the high energy two-particle excitation continua become more localized when $\mathrm{g}=0.2$. A weak signal appears at the energy level around $\omega=3J_1$. In Fig.~\ref{fig:fig5}(n), the plateau structures disappear, except the one contributed by the two-doublon excitations. Fig.~\ref{fig:fig5}(n) tells us that the signal at the energy level around $\omega=3J_1$ is from two-quarton excitations.

In Fig.~\ref{fig:fig5}(o), all six different kinds of two-particle excitation continua become even more localized and the signals are well distinguished. In Fig.~\ref{fig:fig5}(p), no plateau structures are observed in the DOS spectrum. By comparing Fig.~\ref{fig:fig5}(o) and Fig.~\ref{fig:fig5}(p), it can be deduced that the high energy two-particle excitations are spinon-doublon, spinon-quarton, two-doublon, doublon-quarton, and two-quarton excitations while the low energy two-particle excitation is the two-spinon excitation. These two-particle excitations are excited at the energy levels $\omega=J_1$, $\omega=1.5J_1$, $\omega=2J_1$, $\omega=2.5J_1$, and $\omega=3J_1$. According to the calculations in Eqs.~\eqref{eq:rate2ass}~--~\eqref{eq:rate2aqq} and the results stated in Table.~\ref{table:1}, we can conclude that the highest and the second highest transition rate contributions arise from the spinon-doublon and the two-spinon excitations, while the lowest transition rate is from the two-quarton excitation. The transition rate calculations justify the relative RIXS intensity strength in the indirect RIXS spectrum in Fig.~\ref{fig:fig5}(o).

\begin{figure}[t]
\centering
\centerline{\includegraphics[width=8.5cm]{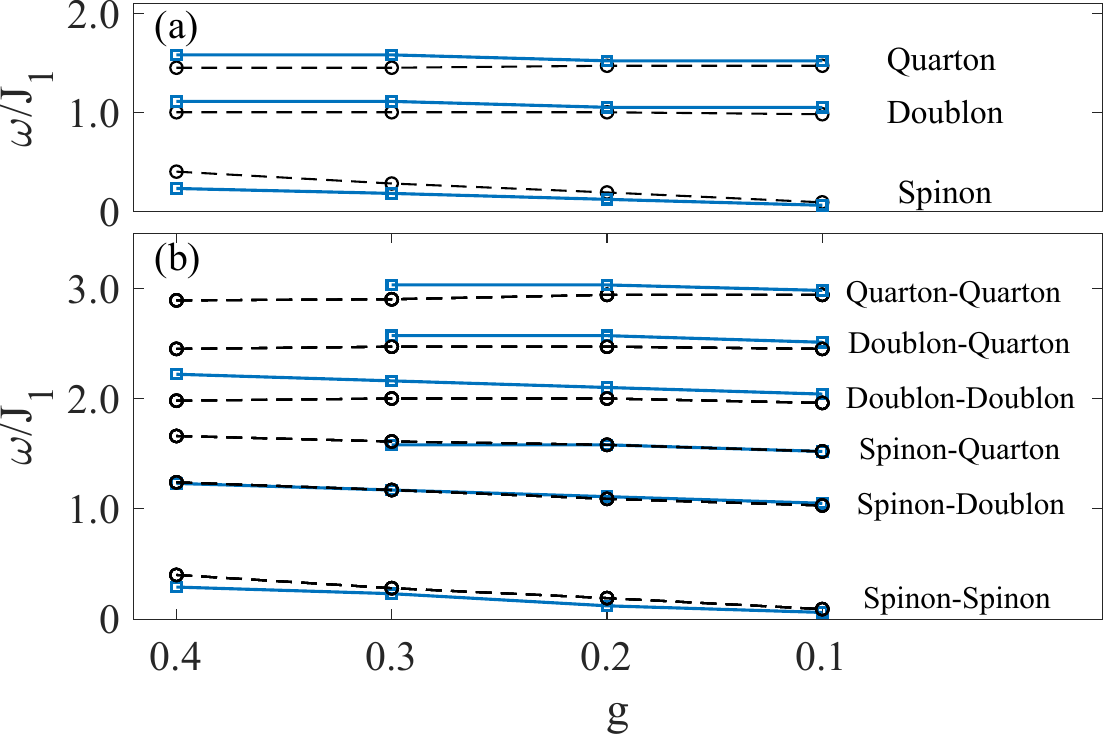}}
\caption{The energy levels for the DOS signals and the integrated RIXS intensity. (a) The dashed lines with black circles are the energy levels for the single-particle DOS and the blue lines with squares show the energy levels for the integrated direct RIXS intensity $\int^{2\pi}_0 S\left(q,\omega\right)dq$. (b) The dashed lines with black circles indicate the energy levels for the two-particle DOS and the blue lines with squares represent the energy levels for the integrated indirect RIXS intensity $\int^{2\pi}_0 O\left(q,\omega\right)dq$.}
\label{fig:fig6}
\end{figure}

To ensure that the DOS results are consistent with the numerical RIXS results obtained by the CV method, we integrate the RIXS spectra over the entire BZ. The integrated RIXS intensity plots are displayed in Fig.~\ref{fig:fig5}. In Fig.~\ref{fig:fig6}, we show the comparison of the numerically computed DOS and the integrated RIXS intensity result for different intertrimer coupling strength g. Both in Fig.~\ref{fig:fig6}(a) (which represents the $L-$ edge) and Fig.~\ref{fig:fig6}(b) (which represents the $K-$ edge), we notice that the difference in the energy levels of the DOS when compared to the integrated RIXS peaks is almost the largest when g=0.4. The numerical difference in the values from the two different calculation approaches diminishes as g decreases. This is because of the dispersion of the energy of the spin excitations which indicates that the relatively stronger intertrimer coupling g may cause deviations between the DMRG results and the perturbative analysis. When g=0.1, slight differences between the peak locations of the DOS signals and the integrated RIXS signals still exist. This is induced by the size effect of the perturbative analysis, where 18 sites (6 trimers) are considered in the calculation, while the DMRG computation includes 60 sites (20 trimers).

\section{Conclusions}\label{sec:conclusions}
We investigated the spin dynamics of the spin-1/2 antiferromagnetic trimer spin chain by calculating both the direct and the indirect RIXS spectrum using the Krylov-space CV method in DMRG simulation. We computed the excitation spectra for both direct and indirect RIXS using $S(q,\omega)$ and $O(q,\omega)$, respectively. Note, the four-spin correlation function $O(q,\omega)$ (which is related to the spin-quadrupolar correlator~\cite{OnishiJPSJ.84.083702}) captures the two-particle excitations, while the two-spin correlation function $S(q,\omega)$ detects the single-particle excitations. Direct RIXS, which probes single-particle excitation, reveals the high energy doublon and the quarton at the energy level of $J_1$ and $1.5J_1$. This finding is consistent with the conclusions of Cheng $et\;al$.~\cite{ChengNPJQM2022}, where QMC simulation was performed to obtain the DSF of the trimer spin chain. Besides single-particle excitation, our calculation suggests that RIXS spectroscopy is able to observe two-particle collective excitations. The low energy excitation creates two spinons propagating along the trimer spin chain in the indirect RIXS process, which is the reason why the low energy two-spinon continua exist in both the direct and the indirect RIXS spectra. Our proposed RIXS mechanism allows us to conclude that the two-particle excitations involving high energy single-particle excitations are capable of supporting the spinon-doublon, the spinon-quarton, the two-doublon, the doublon-quarton, and the two-quarton excitations. By comparing the DOS spectra and the DMRG simulation results, it can be inferred that the resonant energy for each two-particle excitation is the linear combination of the energy for single-particle excitations when g is as small as 0.1. For example, the doublon and the quarton excitations are at the energy level of $\omega=J_1$ and $\omega=1.5J_1$ while the resonant energy for the two-doublon and doublon-quarton excitations are at $\omega=2J_1$ and $\omega=2.5J_1$. All the high energy two-particle excitations have comparable spectral weights. Based on the transition rate calculations, the highest RIXS intensity occurs for the spinon-doublon excitation. The experimentally tested spin-1/2 trimer chain Na$_2$Cu$_3$Ge$_4$O$_{12}$ with intratrimer interaction $J_1=20.6$ meV~\cite{BeraNatComm2022} excites doublons at $\omega=16$ meV and quartons at $\omega=30$ meV. Based our calculations it can be deduced that the high energy two-particle excitations spread in the energy range approximately from $\omega=16$ meV to $\omega=60$ meV. Currently, the RIXS instrumentation resolution for the direct $L$-edge and the indirect $K$-edge are 15 meV and 25 meV, respectively. Thus, RIXS would be capable of experimentally detecting both the single-particle and the two-particle excitations in the spin-1/2 trimer chain.

\begin{acknowledgements}
J.L.L. and D.X.Y. thank Muwei Wu for helpful discussions. T.D. thanks A. Nocera for helpful discussions on DMRG. J.L.L. and D.X.Y. are supported by NKRDPC2022YFA1402802, NSFC-92165204, NSFC-12494591, Leading Talent Program of Guangdong Special Projects (No. 201626003), Guangdong Provincial Key Laboratory of Magnetoelectric Physics and Devices (No. 2022B1212010008), Research Center for Magnetoelectric Physics of Guangdong Province (No. 2024B0303390001), and Guangdong Provincial Quantum Science Strategic Initiative (No. GDZX2401010). T.D. acknowledges the hospitality of KITP at UC-Santa Barbara. A part of this research was completed at KITP and was supported in part by the National Science Foundation under Grant No. NSF PHY-1748958. T.D. acknowledges Augusta University High Performance Computing Services (AUHPCS) for providing computational resources contributing to the results presented in this publication. J.Q.C. is supported by NSFC-12047562, Special Project in Key Areas for Universities in Guangdong Province (No. 2023ZDZX3054) and Dongguan Key Laboratory of Artificial Intelligence Design for Advanced Materials (DKL-AIDAM).
\end{acknowledgements}

\begin{appendix}
\section{Indirect RIXS transition probabilities}\label{app:rateindrixs}
The formulae utilized to compute the intertrimer two-particle excitation transition probabilities are given by
\begin{widetext}
\begin{subequations}
\begin{eqnarray}
\label{eq:rate2ass}
I_{GG\rightarrow GG}&=&\sum_{\sigma=\pm}\sum_{\left(i,\overline{i}\right)}\sum_{n'=1,2}\left|\prescript{i}{n+1}{\bra{0}}\prescript{\overline{i}}{n}{\bra{0}}S^{\sigma}_{n',n'+1}\ket{0}^{i}_n\ket{0}^{\overline{i}}_{n+1}\right|^2, \\
\label{eq:rate2asd}
I_{GG\rightarrow GD}&=&\sum_{\sigma=\pm}\sum_{\left(i,\overline{i}\right)}\sum_{n'=1,2}\left|\prescript{i}{n+1}{\bra{1}}\prescript{\overline{i}}{n}{\bra{0}}S^{\sigma}_{n',n'+1}\ket{0}^{i}_n\ket{0}^{\overline{i}}_{n+1}\right|^2+\left|\prescript{i}{n+1}{\bra{0}}\prescript{\overline{i}}{n}{\bra{1}}S^{\sigma}_{n',n'+1}\ket{0}^{i}_n\ket{0}^{\overline{i}}_{n+1}\right|^2, \\
\label{eq:rate2asq}
I_{GG\rightarrow GQ}&=&\sum_{\sigma=\pm}\sum_{\left(i,\overline{i}\right)}\sum_{\left(i,\overline{j}\right)}\sum_{\left(\overline{i},j\right)}\sum_{n'=1,2}\left|\prescript{j}{n+1}{\bra{2}}\prescript{\overline{i}}{n}{\bra{0}}S^{\sigma}_{n',n'+1}\ket{0}^{i}_n\ket{0}^{\overline{i}}_{n+1}\right|^2+\left|\prescript{i}{n+1}{\bra{0}}\prescript{\overline{j}}{n}{\bra{2}}S^{\sigma}_{n',n'+1}\ket{0}^{i}_n\ket{0}^{\overline{i}}_{n+1}\right|^2, \\
\label{eq:rate2add}
I_{GG\rightarrow DD}&=&\sum_{\sigma=\pm}\sum_{\left(i,\overline{i}\right)}\sum_{n'=1,2}\left|\prescript{i}{n+1}{\bra{1}}\prescript{\overline{i}}{n}{\bra{1}}S^{\sigma}_{n',n'+1}\ket{0}^{i}_n\ket{0}^{\overline{i}}_{n+1}\right|^2, \\
\label{eq:rate2adq}
I_{GG\rightarrow DQ}&=&\sum_{\sigma=\pm}\sum_{\left(i,\overline{i}\right)}\sum_{\left(i,\overline{j}\right)}\sum_{\left(\overline{i},j\right)}\sum_{n'=1,2}\left|\prescript{j}{n+1}{\bra{2}}\prescript{\overline{i}}{n}{\bra{1}}S^{\sigma}_{n',n'+1}\ket{0}^{i}_n\ket{0}^{\overline{i}}_{n+1}\right|^2+\left|\prescript{i}{n+1}{\bra{1}}\prescript{\overline{j}}{n}{\bra{2}}S^{\sigma}_{n',n'+1}\ket{0}^{i}_n\ket{0}^{\overline{i}}_{n+1}\right|^2, \\
\label{eq:rate2aqq}
I_{GG\rightarrow QQ}&=&\sum_{\sigma=\pm}\sum_{\left(i,\overline{i}\right)}\sum_{n'=1,2}\left|\prescript{j}{n+1}{\bra{2}}\prescript{\overline{j}}{n}{\bra{2}}S^{\sigma}_{n',n'+1}\ket{0}^{i}_n\ket{0}^{\overline{i}}_{n+1}\right|^2,
\end{eqnarray}
\end{subequations}
where the combinations of $\left(i,\overline{i}\right)$, $\left(i,\overline{j}\right)$, and $\left(\overline{i},j\right)$ are defined in Sec.~\ref{sec:method}.
\end{widetext}
\end{appendix}
\bibliography{ref}

\end{document}